%% file: CacheEnergySavingTechniqueForMultitaskingSystems.tex
\documentclass[conference,12pt]{IEEEtran}
\ifCLASSINFOpdf
\else
\fi

\usepackage{epsfig}

\usepackage[square, comma, sort&compress,numbers]{natbib}
\usepackage{flushend}

\usepackage{subfigure}
\usepackage{graphicx}
\usepackage{amsmath}
\usepackage[lined,ruled,commentsnumbered,linesnumbered]{algorithm2e}
\usepackage{times}

\newcommand{\Keywords}[1]{\par\noindent
{\small{\em Keywords\/}: #1}}
\usepackage{url}

\makeatletter
\def\url@leostyle{%
  \@ifundefined{selectfont}{\def\UrlFont{\sf}}{\def\UrlFont{\small\ttfamily}}}
\makeatother
\usepackage{setspace}
\begin{document}
\onecolumn

%
\title{ A Cache-Coloring Based Technique for Saving Leakage Energy In Multitasking Systems}

 \author{\IEEEauthorblockN{Sparsh Mittal }
 \IEEEauthorblockA{Department of Electrical and Computer Engineering \\
 Iowa State University, Ames, Iowa 50011, USA\\
  Email: sparsh0mittal@gmail.com}
 }

\maketitle

\input{abstract1}

\IEEEpeerreviewmaketitle

\input{intro1}

\input{motivation1}
\input{background1}

\input{method1}
\input{results1}

\input{conclusion1}



%

\bibliographystyle{IEEEtran}
\bibliography{References}

\end{document}

%% file: abstract1.tex
\begin{abstract}
There has been a significant increase in leakage energy dissipation of CMOS circuits with each technology generation. Further, due to their large size, last level caches (LLCs) spend a large  fraction of their energy in the form of leakage energy and hence, addressing this has become extremely important to meet the challenges of chip power budget.

 For addressing this, several techniques have been proposed. However, most of these techniques require offline profiling and hence cannot be used for real-life systems which usually run multitasking programs, with possible pre-emptions. In this paper, we propose a dynamic profiling based technique for saving cache leakage energy in multitasking systems. Our technique uses a small coloring-based profiling cache, to estimate performance and energy consumption of multiple cache configurations and then selects the best (least-energy) configuration among them. Our technique uses non-intrusive profiling and saves energy despite intra-task and inter-task variations; thus, it is suitable for multitasking systems. Simulations performed using workloads from SPEC2006 suite show the superiority of our technique over an existing cache energy saving technique. With a 2MB baseline cache, the average saving in memory sub-system energy is 22.8\%.


\Keywords{ Multitasking system, Cache Leakage Energy Saving, Energy Efficiency, Dynamic Cache Reconfiguration, }

\end {abstract}

%% file: intro1.tex
\section{Introduction}

Energy efficiency has now become the first-class constraint in the design of computing systems \cite{Mit_DRAMsurvey}. Recently a great deal of research work has focused on reducing leakage energy
in caches \cite{li2002leakage,wang2010leakage,wang2009sacr}. However, the
existing leakage energy saving techniques fail to address the needs of real-life
systems. Real-life systems usually run many applications in a multitasking
manner, where different tasks may be scheduled or preempted in any order
depending on the user requirement (e.g. round robin scheduling
\cite{reddy2010_multitasking_roundrobin}). The cache energy saving techniques
which require static profiling do not scale to these systems due to a large
difference in profiled and actual runs. A few other techniques require intrusive
profiling and hence introduce a large time/space overhead. Moreover, use of
energy saving techniques results in increase in the number of off-chip accesses
which may even offset the energy saved in the cache. Thus, to address the needs
of real-life systems, new leakage energy saving technique are required which
also strike a suitable balance between energy saving and performance loss.
 
 In this paper, we propose a cache-coloring based technique for saving leakage energy in multitasking systems. Our technique uses a
small hardware component called profiling cache which provides
miss rates for multiple cache sizes, including non-power-of-two cache sizes.
Using this, along with the memory stall cycle (MSC) estimation model, our technique
estimates program execution time under multiple possible cache configurations.
Then, for these configurations, memory sub-system energy is estimated and using the energy saving algorithm, the cache is reconfigured to the best (i.e.
least-energy) configuration.
 
 Our technique has several salient features, which make it especially suited to
multitasking systems. By virtue of using dynamic profiling, our technique can easily account for the the effect of
preemption and still save considerable energy without sacrificing performance.
It optimizes for memory sub-system energy, and not merely cache energy and thus, it does not harm other components of the processor. Finally, it optimizes directly for energy and does not work by indirectly (i.e., by controlling other parameters such as  miss-rate).

 Microarchitectural simulations have been performed with Sniper simulator
\cite{CarHei2011_Sniper} and benchmark programs from SPEC2006. Further, our technique has been compared with decay-cache technique.  These experiments show the superiority of our technique in saving leakage energy. The average saving in
memory sub-system energy and EDP, with 2MB baseline cache are 22.8\% and 19.2\%
respectively.

%% file: motivation1.tex


%% file: background1.tex
\section {Background and Related Work}\label{sec:relatedwork}
Modern computing systems execute several resource-intensive applications (e.g. \cite{kumar2011distributed,MitMit2011_QA,PanMit2009_Baywave,jana2013mobile}). To meet the performance demands of these applications, designers provision large amount of on-chip resources. However, in the case when average demand remains low, significant amount of energy is wasted in last level caches in the form of leakage energy. To address these, leakage energy saving techniques are extremely important. 

In literature, several techniques have been proposed for saving cache leakage energy (e.g. \cite{PowSeh00_GatedVdd,li2002leakage,wang2010leakage,mittal2013PhDThesis}). Yang et al. \cite{YanPow02_HybridCache} propose an approach for dynamically reconfiguring the cache. In their technique, a \textit{miss-bound} is calculated for each application using offline analysis. Then, during the actual run, in each interval, the miss rate is compared with the miss-bound and based on it, the cache size is adjusted.
Wang and Mishra \cite{wang2009sacr} propose a method for saving energy in real-time multitasking systems. Their method requires offline computation for storing both performance-optimal and energy-optimal configuration(s) for each task and for each possible execution phase. This information is used during actual run. However, real-life systems execute trillions of instructions of arbitrary applications and for such cases, offline profiling imposes huge overhead and may even become impossible. Further, in the case of multitasking programs with possible pre-emption, due to the large difference between the profiled and actual-runs, the conclusions derived from offline  profiling data could be highly misleading. 
 
Kaxiras et al. \cite{KaxHuz01_CacheDecay} propose decay cache technique for turning off the cache lines which have not been
accessed for a certain number of cycles, called ``decay interval''. However, this technique faces two limitations. Firstly, the techniques based on a fixed decay interval are shown to be less effective for L2 than for L1 \cite{AbeGon05_IATAC}. Secondly, since the optimal value of decay interval changes widely for different programs; the requirement of tuning makes this technique difficult to use in real-life applications \cite{ZhoTob03_AMC}.  
    
 Mittal \cite{mittal2013PhDThesis} uses selective sets and selective ways approach to reconfigure the cache to save leakage energy in LLCs. However, selective sets introduces larger overhead of reconfiguration than the cache-coloring approach. This is due to the fact that on a change in the set-counts, the set-decoding scheme changes and the entire cache needs to flushed. Our coloring scheme reduces this overhead since on a change in the number of active colors, only the contents of the affected cache colors need to be flushed.

%% file: method1.tex
\section{System Implementation}\label{sec:systemimplementation}
Our technique is based on the following observation. 
Since applications show a large intra-application and inter-application variations in their cache requirements, in any interval, a suitable amount of cache can be allocated to an application and the remaining cache can be turned-off for saving leakage energy.

To resize the cache in a dynamic manner, we use cache coloring technique \cite{KesHil92_PageColoring}. Briefly, the cache is divided into multiple non-overlapping bins, called cache-colors.  Cache coloring works by mapping the physical pages having similar color to the cache sets in the same color. To allow flexible cache indexing, we use a small \textit{mapping-table} (MT), which is used to map the physical pages to cache colors. The physical page are grouped into the same number of \textit{regions} as the cache colors  and  the region-to-color mapping is stored in the MT. For an operating page size of 4K and L2 block size of 64, for 4-way, 2MB L2 cache, there are 128 colors and for 8-way, 2MB L2 cache, there are 64 colors. By controlling the number of active cache colors, the amount of cache allocated to the application is controlled. The remaining colors can be turned-off to save cache leakage energy. 

In comparison with page-based mapping design, our design does not have the flexibility to control the mapping of every OS page. Instead, we can only control the mapping of each memory region which contains multiple physical pages. However, we believe that the granularity of the memory region is sufficient enough for the purpose of saving cache energy, as shown through the results.

To estimate the dynamic energy consumption of various possible cache configurations, we use a small microarchitectural component, called profiling cache, while helps in estimating energy value of different configurations. Profiling cache is based on the principle of set sampling \cite{KesHil94_SamplingSetTime}. The profiling cache is a tag-only cache and has similar replacement policy and associativity as the L2 cache. In this paper, we use the sampling-ratio ($R$) of 64, thus, the profiling cache samples only 1 out of 64 sets of the L2 cache.

We further extend it to a multi-level profiling cache, each level profiling a cache size of  $N/16$, $2N/16$, $4N/16$, $8N/16$, $12N/16$, $16N/16$, where $N$ shows the number of L2 colors (or equivalently L2 cache size). We refer to each cache size profiled as a \textit{profiling point}. A unique feature of our profiling cache is its ability to profile caches whose set-counts are \emph{not} power-of-two values. This is achieved by using cache coloring scheme. This is a significant improvement over previous works based on cache reconfiguration. We explain the reasoning behind the choice of these six cache sizes in Section~\ref{sec:makealgorobust}.   
  
If the number of sets in L2 cache are $P$, then the total number of sets in  multi-level profiling cache ($S$) can be calculated as follows 
\begin{equation}\label{eq:sumofprof}
S = \dfrac{P}{16R} + \dfrac{2P}{16R} + \dfrac{4P}{16R} + \dfrac{8P}{16R} +\dfrac{12P}{16R} + \dfrac{16P}{16R} = \dfrac{43P}{16R} 
\end{equation}

We now compute the overhead of profiling cache ($F_{prof}$) compared to L2 cache size. For this, we assume a $W$ way L2 cache, with $L$ byte block size and $T$ bit tag. Thus,
\begin{equation}
F_{prof} = \dfrac{\text{ProfilingCacheSize}}{\text{L2CacheSize}}=\dfrac{S\times W \times T }{P\times W\times(8L+T)}  
\end{equation}
\begin{equation}
\implies F_{prof} = \dfrac{43T}{16R(8L+T)}
\end{equation}

For $R=$64, $T=$40, $L=$64, we get $F_{prof}=$.003 or 0.3\%. Thus, profiling cache has a small overhead compared to L2 cache. Henceforth, we use the word profiling cache to indicate a multi-level profiling cache, unless otherwise mentioned. 

 

\section{Predicting Memory Stall Cycle For Energy Estimation} \label{sec:StallCycle}
In addition to dynamic energy, we also account for leakage energy of memory sub-system under different L2 configurations. For this purpose, program execution time under those configurations needs to be estimated. To address this, we use  hardware counters for continuously measuring effective memory stall cycles, taking into account possible overlap with other miss events (e.g. branch misprediction, L1 miss). Thus, the total-cycles of the program is decomposed into base-cycles and stall-cycles.  Further, extra counters are used with profiling cache for also measuring the number of L2 load misses under different cache configurations. 

We assume that in an interval $i$ with configuration $C_{\star}$, the effective stall-cycles (StallCycles$_i(C_{\star})$) are proportional to the number of load-misses (LoadMisses$_i(C_{\star}$)). Thus, their ratio (termed as penalty-per-miss or PPM$_i$) is independent of the number of load-misses themselves. Using this, the StallCycles$_i(C)$ for any configuration $C$ can be estimated by multiplying PPM$_i$ with the number of load-misses under that configuration. Further, from StallCycles$_i(C)$ value, the total-cycles (or equivalently execution time) under configuration $C$ are computed by adding base-cycles value to it. Using this, the leakage energy of the program under any configuration can be estimated, as shown in Section \ref{sec:energymodel}.

\section{ Energy Saving  Algorithm}\label{sec:makealgorobust}
We now show our energy saving algorithm, which works in each interval. It intelligently selects a small number of candidate configurations in each interval, estimates their energy and then selects the most energy efficient configuration from the candidate configurations.  We now  discuss the working of energy saving algorithm in detail. 

Let ConfigSpace and $D$ to show the set of candidate configurations and its cardinality (i.e. the number of candidate configurations) respectively.  To keep the reconfiguration overhead small and avoid oscillation, the energy saving algorithm selects configurations in neighborhood of $C_{\star}$ using following criterion. Firstly, $D$ is set to a small value. In this paper, $D$ is  taken as 11  which includes $C_{\star}$ itself. Secondly, to avoid any thrashing, the energy saving algorithm  selects only those configurations which have at-least $Min$ active colors. In this paper, $Min$ is set to $N/16$. Further, the granularity of cache allocation is taken as two colors. This enables testing a wider range of configurations, while still keeping algorithm overhead small.

 Finally, our technique computes \textit{marginal gain} values and utilizes them to make an intelligent guess about the candidate configurations.  At any color value $C$, the value of marginal gain, $G(C)$, is defined as the reduction in cache misses on increasing a single color. Thus, $G(C)$ is a measure of utility of increasing unit cache resource of the program. Consistent with the previous works, we assume that between two profiling points, the  number of misses vary linearly with cache size and hence, the marginal gain remains constant.  Intuitively, if a program has low $G(C_{\star})$ value, then the configurations with smaller cache size are likely to be energy efficient and vice-versa. Using this simple observation, the energy saving algorithm selects candidate configurations such that if $G(C_{\star}) \le \lambda$,  ConfigSpace contains maximum 6 valid configurations smaller than $C_{\star}$ and maximum 4 `valid' configurations larger than $C_{\star}$. Here $\lambda$ is a threshold and is set to $200$. Similarly, if $G(C_{\star}) > \lambda$,  ConfigSpace contains maximum 4 valid configurations smaller than $C_{\star}$ and maximum 6 valid configurations larger than $C_{\star}$. As an example, when $\lambda$=$200$, $N$=$64$, $C_{\star}$=$40$, $G(C_{\star})$=$150$, then ConfigSpace=$\{ 28,30,32,34,36,38,40,42,44,46,48\}$. If $G(C_{\star})$ were to be $250$, then ConfigSpace=$\{ 32,34,36,38,40,42,44,46,48,50,52\}$.

\section{Implementation}
To switch off a cache line we use the well-known NMOS gated-$V_{dd}$ technique with dual Vt, wide, with charge pump \cite{PowSeh00_GatedVdd}. This technique results in minimal impact on access latency but introduces a 5\% area penalty. We account for the effect of this on leakage, as shown in Section \ref{sec:simulationmethodology}. 
  
The reconfigurations are handled as follows. When the number of active cache colors is decreased, the contents of the disabled cache colors are flushed. Also, the memory regions mapped to these colors are remapped to other active colors. On the other hand, when the number of active cache colors are increased, some memory regions, which were mapped to another color, are remapped to newly active colors and the blocks of those memory regions in their previous (original) colors are flushed. Our reconfiguration scheme  incurs one-time overhead, but is simple and requires less state storage than previous schemes \cite{RanAdv00_Recon}. Since reconfigurations only happen at the end of an interval, thus, cache block turning off/on  does not lie at critical path of cache access. Finally, since algorithm uses a large interval length (Section \ref{sec:simulationmethodology}), the overhead of reconfigurations is amortized over the phase length.

 
%
%
%
%
%
%
%
%
%
%
%

%

%% file: results1.tex
\section{Simulation Methodology } \label{sec:simulationmethodology}
\subsection{Platform and Workload}
To perform microarchitectural evaluation we have used Sniper \cite{CarHei2011_Sniper}, a Pin based state-of-art simulator, which has been validated against real hardware \cite{CarHei2011_Sniper}. We model a 1.5 GHz, 4 wide processor with ROB size of 128. L1 data/instruction caches are 32KB, 4 way, LRU, 64B line size and have a latency of 6ns. The unified L2 is 2MB, 8 way, 64B line size LRU with 12ns latency. The main memory has a latency of 60ns. The memory queue has a peak bandwidth of 6GB and the queue contention is also modeled.

Follwing Phansalkar et al. \cite{phansalkar2007subsetting}, we use  12 benchmarks from SPEC2006, which represent the behavior of entire SPEC2006 suite. This helps us in simulating the representative behavior of SPEC2006 suite, while still limiting the simulation time.   These benchmarks are 6 each from floating point (cactusADM,  lbm, milc, povray, soplex, wrf) and  integer point (gcc, hmmer, libquantum, mcf, sjeng, xalancbmk) benchmarks. 

To construct 3-paired task-sets from these benchmarks, we proceed as follows (Table \ref{tab:workloads}). Benchmarks were arranged in alphabetical order. Then, MW1-MW4, MW5-MW8 and MW9-MW12 were made by pairing benchmarks at a modulo distance (i.e. circular order) of 1, 3 and 4. Each benchmark was fast forwarded for 10B instructions and each benchmark in the task set of three benchmarks executes 300M instructions.

\begin{table}[htbp]
  \centering
  \caption{Multitasking workloads }
    \begin{tabular}{|c|c|c|c|} \hline
    MW1       &  cactus gcc hmmer & MW7       &  povray wrf hmmer \\\hline
    MW2       &  lbm libquant mcf & MW8       &  mcf sjeng xalan \\\hline
    MW3       &  milc povray sjeng & MW9       &  cactus libquant sjeng \\\hline
    MW4       &  soplex wrf xalan  & MW10      &  gcc mcf soplex \\\hline
    MW5       &  cactus lbm milc & MW11      &  hmmer milc wrf \\\hline
    MW6       &  soplex gcc libquant & MW12      &  lbm povray xalan \\\hline

    \end{tabular}%
   \\ Here cactus=cactusADM, libquant=libquantum, xalan= xalancbmk
  \label{tab:workloads}%
\end{table}%

\begin{figure}[htp]
 \centering
  \includegraphics [scale=0.50] {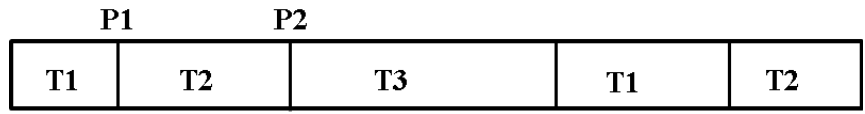}
 \caption{Task model }\label{fig:taskmodel}
 \end{figure}

 Our task set model is shown in Figure \ref{fig:taskmodel}. First T1 is scheduled and then at point P1 of its execution, T2 arrives, which preempts T1 and begins execution. At point P2 of the execution of T2, T3 arrives, which preempts T2 and begins execution. Once T3 is finished, T1 is scheduled and then T2 is scheduled. For simulation purpose, we arbitrarily chose P1 and P2 values to be 80M and 130M instructions of T1 and T2 respectively, for all task sets (although any other preemption points would work fine also). Each application (task) has its own separate physical memory address space, and thus, they do not share data.

\subsection{Comparison with Decay Cache Technique}
We compare our technique with a well-known cache energy saving, namely decay cache technique (DCT) \cite{KaxHuz01_CacheDecay}. DCT works on the principle that, a line which has not been accessed for the duration of `decay interval' (DI) can be deactivated, since its usefulness has decayed \cite{KaxHuz01_CacheDecay}. We have chosen DCT for two main reasons. Firstly, DCT, like our technique, is a state destroying technique and hence allows a fair comparison. Secondly, it is a well-known technique and has been used by other researchers also (e.g. \cite{HanHri02_TVLSI,AbeGon05_IATAC,li2002leakage,HanHri02_TVLSI}).  To achieve largest possible energy saving, both data and tags are decayed. 

Regarding the choice of the decay interval (DI), Kaxiras et al. suggest using large DI for L2 (100 times that of L1 caches), such as 6,400K (6.4M) or 12.8M cycles \cite{KaxHuz01_CacheDecay}. To find a good DI, we conducted experiments with DI of 3.2M, 6.4M, 12.8M and 19.2M cycles using single-tasking workloads and found that, on average, the decay interval of 6.4M gives the  best (i.e. minimum) EDP results for most workloads. Hence, we choose 6.4M as the decay interval for DCT. 

When a change in running task happens; for DCT, decay counters \cite{KaxHuz01_CacheDecay} of all cache blocks are reset (i.e. brought to initial state) and for our technique, profiling cache counters are reset.  Further, for both DCT and our technique, at the time of change in running task, the active cache size is \textit{not} altered. In other words, the active cache size is only altered by the respective techniques and not due to change in running task.

\subsection{Energy Model}\label{sec:energymodel}
We model the energy spent in L2 cache ($E_{L2}$), main memory ($E_{mem}$) and in execution of the algorithm ($E_{Algo}$). Assuming that the energy spent in L2 and memory is composed of both leakage and dynamic energy, we have

\begin{equation}\label{eq:Etot}Energy= E_{L2}+E_{mem}+E_{Algo}\end{equation}

To compute $E_{L2}$, we note that the leakage energy is proportional to the active ratio of the cache \cite{YanPow01_IcacheResize,li2002leakage}. We assume that an L2 miss consumes twice the energy of an L2 hit \cite{HanHri02_TVLSI,mittal2013PhDThesis}.   Thus,
\begin{equation}E_{L2}= E^{dyn}_{L2}\times(H_{L2}+2M_{L2}) + (P^{leak}_{L2}\times Time\times C_{\star})/N\end{equation}
 
Here $P^{leak}_{L2}$ and $E^{dyn}_{L2}$ show the dynamic energy per L2 access and L2 leakage energy per second. Also,  for any interval, we have corresponding $H_{L2}$ = L2 hits, $M_{L2}$ = L2 misses, $Time=\text{Time consumed}$ and $C_{\star}$ = active colors.  We use CACTI 5.3 \cite{cacti_53} to compute the energy values for 8-bank, 8-way caches with 64 byte block size. For 2MB cache, we get the following values: $E^{dyn}_{L2}$ = 1.086 nJ/access and $P^{leak}_{L2}$ = 2.016 Watt. To account for the effect of extra area on leakage energy dissipation, we assume 5\% higher value of $P^{leak}_{L2}$ for both our technique and DCT, but not for baseline LRU cache. 

To compute $E_{mem}$, we take $P^{leak}_{mem}$= 0.18 Watt and $E^{dyn}_{mem}$= 70 nJ  \cite{mittal2013PhDThesis}. Thus, we get \begin{equation}E_{mem}= E^{dyn}_{mem}\times A_{mem}+ P^{leak}_{mem}\times Time \end{equation}

where $A_{mem}$ denotes the number of memory accesses.

The overheads of  block transitions (for both our technique and DCT) and profiling cache (for our technique) are calculated as follows. 
\begin{equation}\label{eq:Ealgo}E_{Algo}= E_{dyn}^{prof}\times A_{prof} +P_{leak}^{prof}\times Time + E_{Tran}\end{equation}
Here $A_{prof}$ shows the number of profiling cache accesses. Further, $E_{dyn}^{prof}$ and $P_{leak}^{prof}$ are dynamic energy per access and leakage power for profiling cache.

We use CACTI along with Eq.~\ref{eq:sumofprof} to calculate energy values for profiling cache. We take an upper bound as $S=64P/16R$ since CACTI only provides values for power-of-two size caches. We compute the energy values for profiling cache, by only taking tag energy values since profiling cache is a tag-only cache. For a profiling cache corresponding to 2MB cache, we get the following values:  $E^{dyn}_{prof}$=0.005nJ/access and   $P^{leak}_{prof}$ = 0.007 Watt.  It is clear that profiling cache consumes a negligibly small fraction of energy compared to the energy consumed by L2 cache.

Assuming that each block-transition takes $0.002$ nJ,  the total energy spent in block transitions is
\begin{equation}E_{Tran}= 0.002\times Q \text{  nJ}\end{equation}
where $Q$ denotes the total number of blocks transitions.

We ignore the overhead of counters, since many processors already contain counters for measuring performance etc. \cite{KaxHuz01_CacheDecay}.

\section{Results on Energy Saving}\label{sec:results}

\begin{figure*}[t]
 \centering
  \includegraphics [scale=0.65] {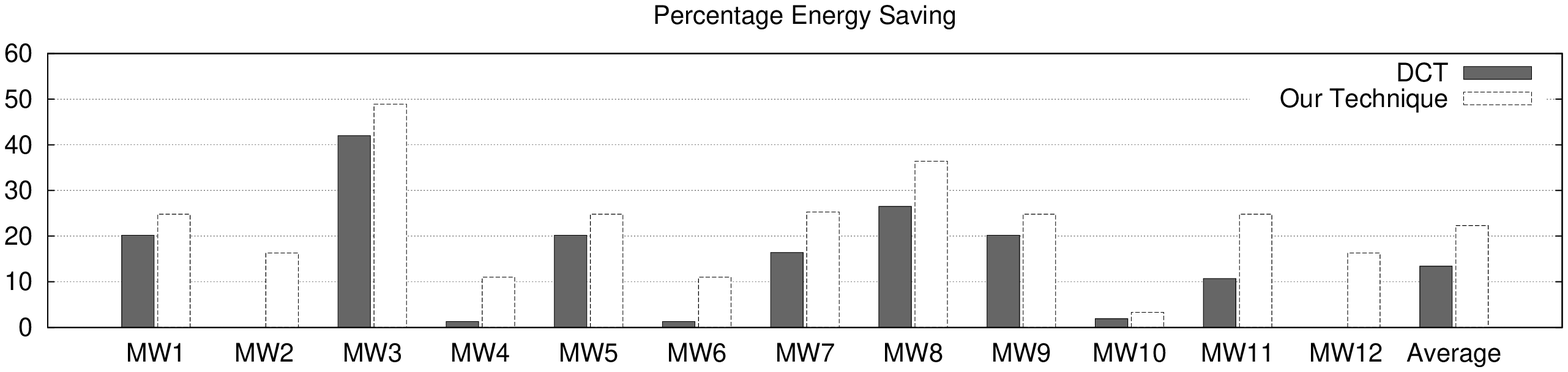}
  \includegraphics [scale=0.65] {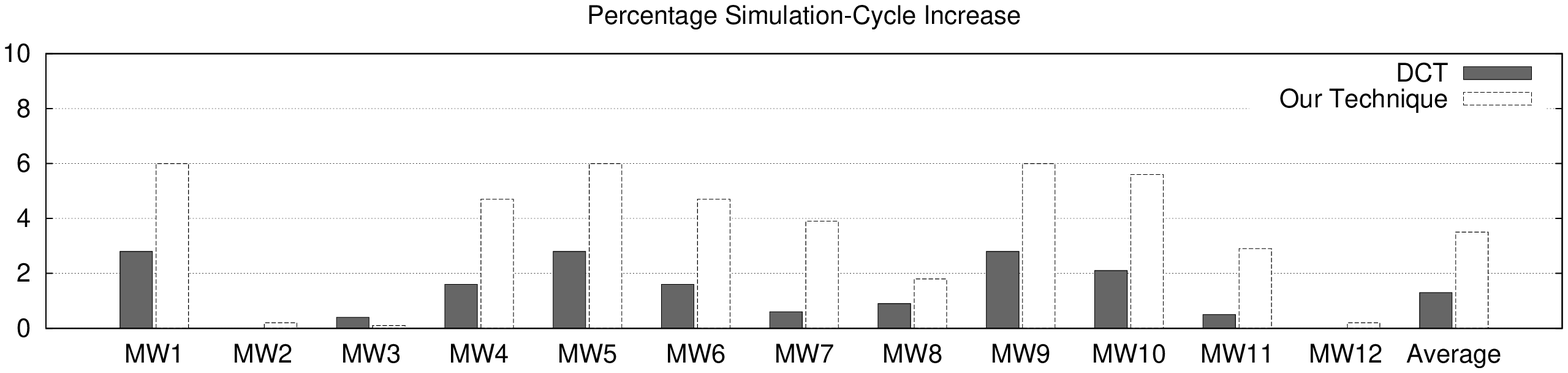}
\caption{Experimental Results for 2MB cache with DCT and our technique }
 \label{fig:exptresult2MB}
 \end{figure*}

Figure~\ref{fig:exptresult2MB}  shows the percentage energy saving over a baseline LRU-managed cache for DCT and our technique. The average saving in energy for DCT and our technique are 15.9\% and 22.8\% respectively. Clearly, our technique provides significantly larger energy saving compared to DCT.

DCT takes into account the number of accesses to a block in a given time and uses this information to turn-off a block. However, if the miss-rate of the program is high, the cache reuse becomes small and hence, the effectiveness of DCT is reduced. In contrast, our technique turns off cache based on marginal gain from allocation of cache. Thus, our technique can detect low cache reuse and aggressively turn-off cache to save energy for such programs. Also, finding a suitable decay interval for a program with DCT requires offline profiling. Our technique does not suffer from this limitation and thus, it can work for a wide variety of benchmarks. 

The advantage of DCT is that it can turn-off cache at very fine granularity and hence, it can potentially save larger energy for some benchmarks. Also, it does not change the set-decoding on reconfiguration and thus does not incur reconfiguration overhead. However, as shown by  energy saving results, it is clear that the reconfiguration overhead of our technique is small enough. Also, for the programs tested, our technique can save large amount of energy and the reconfiguration granularity chosen by our technique is sufficient.

Figure~\ref{fig:exptresult2MB} also shows the percentage increase in simulation time. For DCT the increase in simulation cycles is 1.3\%, while for our technique, this value is 3.5\%. Thus, the increase in simulation cycle with our technique is slightly higher. Still, note that the difference is small and the large amount of energy saving can easily offset the extra energy spent due to increased simulation time. 

The results show the effectiveness of our technique for multitasking systems. The dynamic profiling approach used in our technique is very useful for real-world systems which execute trillions of instructions and run arbitrary applications. Also, our technique can easily take into account the energy consumption of other components of the processor, such as core, peripherals etc. and use that value to guide the algorithm.

%
%
%


%% file: conclusion1.tex
\section{Conclusion}\label{sec:conclusion}
As the amount of data processed by modern computing systems increases \cite{pande2007network,sood2006novel,pande2013embedded}, the size of on-chip caches is also on rise and hence, their energy consumption is becoming a significant fraction of processor energy consumption. We have presented a cache coloring based technique for saving leakage energy in multitasking systems. Our technique uses dynamic profiling with dynamic cache reconfiguration for optimizing memory subsystem energy efficiency.  Most modern major operating systems, such as Linux and other Unix-like systems, Mac OS X etc.  employ pre-emptive multitasking and by its virtue of saving leakage under intra-task and inter-task variations, our technique can serve as a valuable tool for such systems.  Our future work will focus on evaluating our technique in the context of multicore systems.